\documentclass[reprint, superscriptaddress, secnumarabic, amssymb, nobibnotes, aps, prl]{revtex4-1}

\usepackage{graphicx}
\usepackage{epstopdf}
\usepackage[T1]{fontenc}
\usepackage{amsbsy}
\usepackage{gensymb}
\usepackage[T1]{fontenc}
\usepackage{amsmath}
\usepackage{amssymb}
\usepackage{bbm}
\usepackage{braket}
\usepackage{xcolor}
\allowdisplaybreaks
\usepackage{graphicx}
\usepackage[colorlinks=true]{hyperref}  
\hypersetup{
    bookmarks=true,         % show bookmarks bar?
    unicode=false,          % non-Latin characters 
    pdftoolbar=true,        % show Acrobat
    pdfmenubar=true,        % show Acrobat 
    pdffitwindow=false,     % window fit to page when opened
    pdfstartview={FitH},    % fits the width of the page to the window
    pdftitle={XOs$_{2}$},    % title
    pdfauthor={PK Meena},      %author
    pdfsubject={},   % subject of the document
    pdfcreator={},   % creator of the document
    pdfproducer={}, % producer of the document
    pdfkeywords={} {} {}, % list of keywords
    pdfnewwindow=true,      % links in new window
    colorlinks=true,       % false: boxed links; true: colored links
    linkcolor=blue, %red,          % color of internal links (change box color with linkbordercolor)
    citecolor=blue,        % color of links to bibliography
    filecolor=magenta,      % color of file links
    urlcolor=blue           % color of external links
} 

\usepackage[normalem]{ulem}

\newcommand{\figref}[1]{Fig.~\ref{#1}}

\newcommand{\tableref}[1]{Table~\ref{#1}}

\begin{document}

\title{\textrm{Superconductivity in Breathing Kagome-Structured C14 Laves Phase XOs$_{2}$(X = Zr, Hf)}}

\author{P. K. Meena}
\affiliation{Department of Physics, Indian Institute of Science Education and Research Bhopal, Bhopal, 462066, India}
\author{M. Mandal}
\affiliation{Department of Physics, Indian Institute of Science Education and Research Bhopal, Bhopal, 462066, India}
\author{P. Manna}
\affiliation{Department of Physics, Indian Institute of Science Education and Research Bhopal, Bhopal, 462066, India}
\author{S. Srivastava}
\affiliation{Department of Physics, Indian Institute of Science Education and Research Bhopal, Bhopal, 462066, India}
\author{S. Sharma}
\affiliation{Department of Physics, Indian Institute of Science Education and Research Bhopal, Bhopal, 462066, India}
\author{P. Mishra}
\affiliation{Department of Physics, Indian Institute of Science Education and Research Bhopal, Bhopal, 462066, India}
\author{R.~P.~Singh}
\email[]{rpsingh@iiserb.ac.in}
\affiliation{Department of Physics, Indian Institute of Science Education and Research Bhopal, Bhopal, 462066, India}

\begin{abstract}
Recently, the emergence of superconductivity in kagome metals has generated significant interest due to its interaction with flat bands and topological electronic states, which exhibit a range of unusual quantum characteristics. This study thoroughly investigates largely unexplored breathing Kagome structure C14 laves phase compounds XOs$_{2}$ (X = Zr, Hf) by XRD, electrical transport, magnetization, and specific heat measurements. Our analyses confirm the presence of the MgZn$_{2}$-type structure in ZrOs$_{2}$ and HfOs$_{2}$ compounds, exhibiting type-II superconductivity with critical temperature (T$_{C}$) values of 2.90(3) K and 2.69(6) K, respectively. Furthermore, specific heat measurements and an electron-phonon coupling constant suggest the presence of weakly coupled BCS superconductivity in both compounds.
\end{abstract}
\maketitle

\section{INTRODUCTION}
The Kagome lattice, a pivotal platform in condensed matter physics, explores the interplay of geometry, topology, and magnetism via its 2D network of connected triangles with S = 1/2 ions and nearest-neighbor interactions \cite{Hasan, Review, magneticfrustration1, magneticfrustration2}. They exhibit magnetic skyrmions, chiral spin structures, frustration-driven spin-liquid states, and unique quantum features, attracting scientific interest by revealing the Hall effect and time-reversal symmetry breaking (TRSB) in materials \cite{Co3Sn2S2-1, GFS1, GFS2, AV3Sb5-1, AV3Sb5-2, AV3Sb5-3}. Dirac, flat bands, and van Hove singularities in the density of states occurring in the isolated Kagome 3-orbital tight-binding model are nontrivial features. Destructive interference-induced electron wavefunction localization on the Kagome lattice results in a flat band that marks the instability of the system towards crystal lattice distortion or magnetism \cite{Kagome-nontrivial, YMn6Sn6}. The next-nearest neighbor interactions in crystalline solids dispersed the flat band on the Kagome lattice, affecting the density of states and giving rise to a peak commonly associated with superconducting transitions or ferromagnetism \cite{Flatband1, Flatband2, Flatband3, Flatband4}. 

Recently, symmetry breaking in kagome compounds, notably through distortion (breathing), has resulted in the possibility of a higher-order topological phase with quantum behavior comparable to ideal kagome compounds, which has piqued the curiosity of researchers \cite{BreathingKagome-nontrivial1, BreathingKagome-nontrivial2, BreathingKagome-nontrivial3, BreathingKagome-nontrivial4}. The curious superconducting pairing mechanism in breathing kagome compounds hints at unconventional pairing through the interplay of geometric frustration-induced nontrivial states and superconductivity \cite{Kagome-muon1, Kagome-muon2, Kagome-Review}. A recent study on some compounds with a breathing kagome structure \cite{Mg2Ir3Si, BreathingKagome, ROs2} and isostructural Re$_{2}$Hf displaying unconventional superconductivity with TRSB has drawn attention to superconductivity \cite{Re2Hf}. Furthermore, distorted kagome variants in RT$_{3}$X$_{2}$ systems (R = Y, La, Lu, Th, and T = Ru, Os, Rh, Ir, X = B, Si, Ga) exhibit non-trivial band structures and nodeless moderate coupling superconductivity, showing promise for enhanced superconductivity \cite{LaRu3Si2, LaIr3Ga2, YRu3Si2, LaRh3B2, BaPt3B2, CeOs3B2}. Therefore, exploring the superconductivity of more compounds in this distorted structure is essential for further studies on the topic of these unconventional features.

The hexagonal C14 Laves phases XOs$_{2}$ (X = Zr, Hf), characterized by AB$_{2}$ intermetallic compound structure, feature a breathing kagome layer of Os atoms that form a nearly ideal 2D kagome structure with minimal length variation in the Os-Os bonds. These phases have non-symmorphic space group symmetry, with compact Frank-Kasper configurations of tetrahedrally coordinated B atoms around A centers with 1.06$-$1.67 atomic ratio \cite{Laves-Phase1, electronegativity-1, 1, Frank-Kasper, Non-symmorphic1}. The initial band structure calculation revealed the enforced semimetal nature of XOs$_{2}$ with a possible topological nature \cite{TSM}. First-principles calculations indicate a strong metallic nature through a high electronic density of states at the Fermi level \cite{Os2Zr, Pr-ZrOs2, Theortical-ZrOs2, Hf-DFTcalculation}. Additionally, XOs$_{2}$ offer a unique opportunity to study the effects of spin-orbit coupling ($\propto$ Z$^{4}$) on its potential topological and superconducting properties \cite{Topological-SC, Unconventional-SCs}. Despite these insights, the investigation of bulk superconductivity in XOs$_{2}$ (X = Zr/Hf) remains largely unexplored.
\begin{figure*}
\includegraphics[width=2.05\columnwidth]{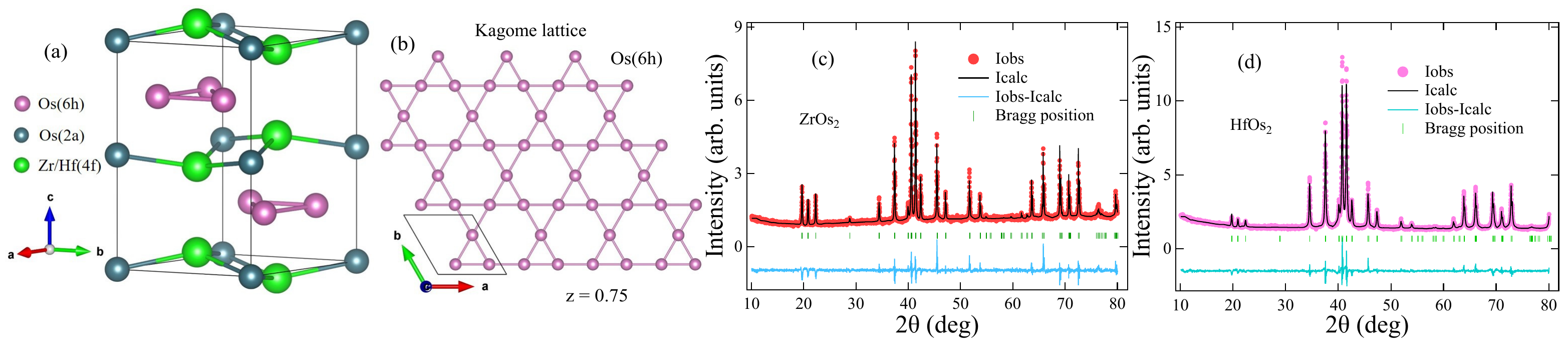}
\caption {\label{Fig1} (a) Crystal structure of XOs$_{2}$ (X = Zr, Hf). The green, olive, and pink spheres represent the X, Os1, and Os2 atoms. (b) The breathing kagome lattice of Os (6h) atoms in the $ab$ plane. Powder XRD pattern with Rietveld refinement for (c) ZrOs$_{2}$. (d) HfOs$_{2}$. Marks, lines, and vertical bars, in that order, show experimental data, theoretical refinement, and the Bragg position. The line at the bottom shows the difference between the observed and calculated data.}
\end{figure*} 

In this paper, we present a detailed investigation of the superconducting properties of the Laves phase C14 XOs$_{2}$ (X = Zr, Hf), featuring a breathing kagome structure through resistivity, magnetization and specific heat measurements. All measurements confirmed fully gapped type II weakly coupled bulk superconductivity with a transition temperature T$_{C}$ of 2.90(3) and 2.69(6) K for ZrOs$_{2}$ and HfOs$_{2}$.

\section{EXPERIMENTAL DETAILS}
Polycrystalline XOs$_{2}$ (X = Zr, Hf) were synthesized using the standard arc melting technique. High-purity ($4N$) elements Zr or Hf, and Os were placed in a stoichiometric ratio of 1:2 in a Cu hearth cooled by water in a high-purity argon ($4N$) environment and melted/flipped several times to improve homogeneity. Powder X-ray diffraction (XRD) using CuK${\alpha}$ radiation with $\lambda$ = 1.54056 $\text{\AA}$ was used to confirm the phase purity and crystal structure. Magnetization measurements were conducted on a Quantum Design Magnetic Property Measurement System (MPMS-7T). Electrical resistivity and specific heat measurements were performed using the Quantum Design Physical Property Measurement System (PPMS-9T).

\section{RESULTS AND DISCUSSION}
\subsection{Sample characterization}
The crystal structure of XOs$_{2}$ (X = Zr, Hf) is shown in \figref{Fig1}(a). It is similar to the structure of MgZn$_{2}$ and is called the C14 Laves phase. Green, olive, and pink balls represent X ($4f$ site), Os1 ($2a$ site) and Os2 ($6h$ site), respectively, constructed by VESTA software \cite{VESTA}. This structure features distinct stacking sequences of triangular nets (X/Os1 atoms), with X atoms occupying interstitial positions slightly offset from Os1 and Kagome nets of Os2 atoms along the c-axis \cite{Lavesphase6, Laves-phase2}. Each unit cell exhibits two vertically stacked kagome layers at $z$ = 0.25 and 0.75 along the $c$ direction with mutual shifts and connections through Os1 atoms. The Os2 atoms create a trigonally distorted kagome lattice where two triangles and two hexagons meet at each net vertex, exhibiting a nearly ideal 2D kagome structure with minimal length variation in Os-Os bonds within the $ab$ plane, as shown in \figref{Fig1}(b). The difference in the Os-Os bond in this layer is $d_{Os-Os}$ ( = 2.64732 and 2.55268 $\text{\AA}$). Rietveld refined the powder XRD patterns of XOs$_{2}$ using the FullProf Suite software as shown in \figref{Fig1}(c) and (d) \cite{FullProf}, confirming their crystallization in hexagonal centrosymmetric structure with the space group P6$_{3}$/$mmc$ (no. 194, $D_{6h}$). The lattice parameters of ZrOs$_{2}$ [$a$ = $b$ = 5.207(9) $\text{\AA}$, $c$ = 8.526(5) $\text{\AA}$ with $V_{cell}$ = 200.27(9) $\text{\AA}^{3}$] and HfOs$_{2}$ [$a$ = $b$ = 5.189(4) $\text{\AA}$, $c$ = 8.487(6) $\text{\AA}$ with $V_{cell}$ = 197.94(7) $\text{\AA}^{3}$], obtained in this work are consistent with the previous report \cite{ZrOs2, Os2Zr}. The variations in lattice constants correspond to the variations in atom size in the respective alloys. Notably, the atomic radii ratio, ideally 1.225, critically influences the formation of the Laves phase structure, playing a pivotal role in the structural characteristics and stability of these alloys \cite{1}. The calculated atomic-radii ratio for both compounds closely matches the ideal value, where the atomic radius is taken from ref. \cite{atomicradius}. The elemental composition determined by the EDS spectrum for both polycrystalline compounds suggests a nominal composition.

\subsection{Superconducting and normal state properties}
\subsubsection{Electrical resistivity}
\begin{figure}
\includegraphics[width=1.0\columnwidth]{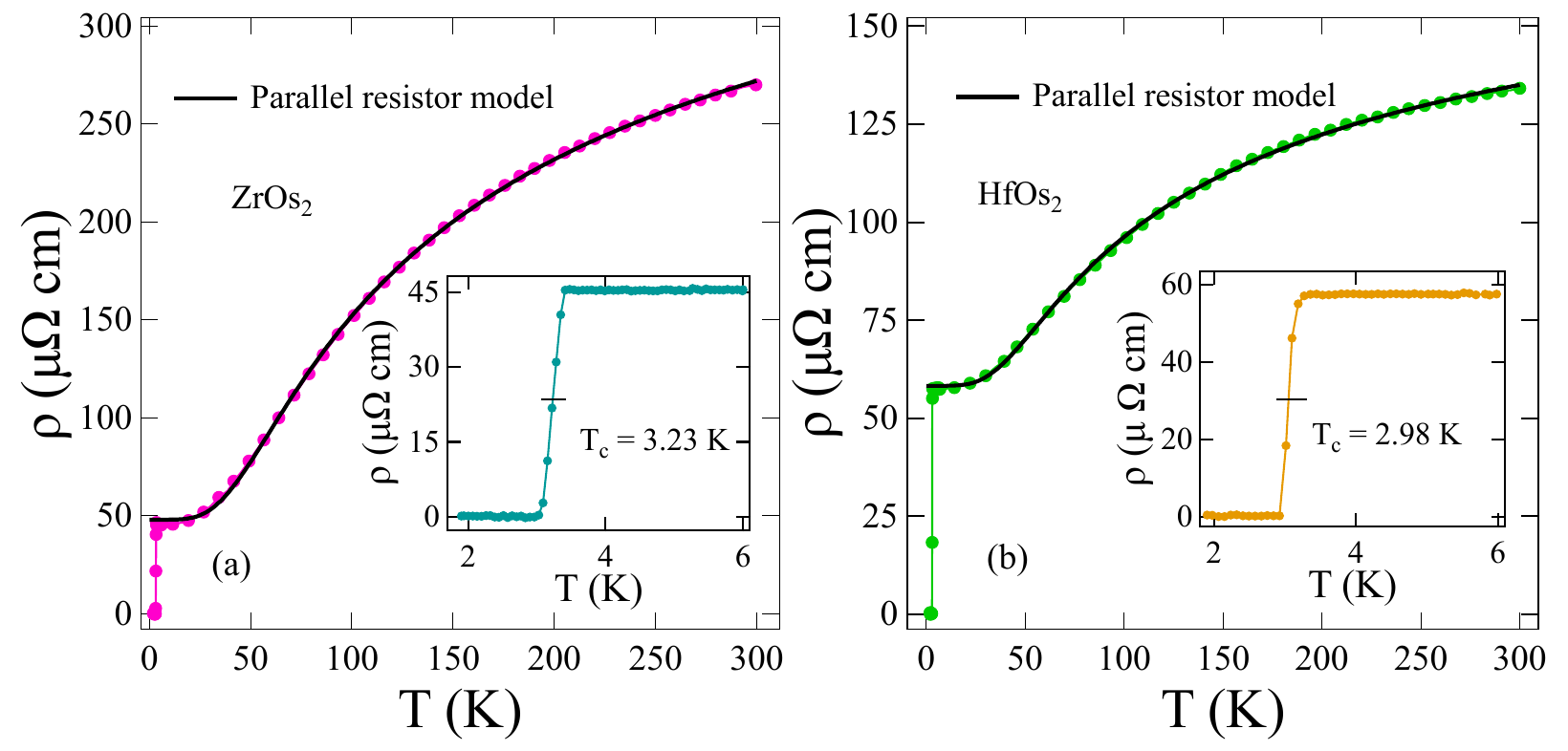}
\caption {\label{Fig2} The resistivity versus temperature curve was fitted using the parallel resistor method for (a) ZrOs$_{2}$ and (b) HfOs$_{2}$. The superconducting transition is confirmed at 3.23 K for ZrOs$_{2}$ and 2.98 K for HfOs$_{2}$, which can be seen in the inset of figures (a) and (b).}
\end{figure}

Temperature-dependent resistivity at zero magnetic field is shown in \figref{Fig2}(a) and (b) for XOs$_{2}$ (X = Zr, Hf), respectively. In the normal state, the resistivity decreases as the temperature decreases, indicating the metallic nature of the compounds. The residual resistivity ratio (RRR), denoted $\rho(300)$ /$\rho(10)$, was obtained as 5.81 and 2.33 for the variants Zr and Hf. A sudden drop in resistivity from the normal state is observed, which ensures a superconducting transition at T$_{C}$ = 3.23 K for ZrOs$_{2}$ and 2.98 K for HfOs$_{2}$. The inset of \figref{Fig2}(a) and (b) represents an expanded resistivity plot, showing the superconducting transition. The measured superconducting transition temperature is consistent with previous reports \cite{LS1, Os2Zr}. The conservatively determined values obtained are higher than those reported for some cubic Laves phases with C14 Laves phases based on Ir and Ru \cite{Ir-LavesPhase, Ru-LavesPhase}. The normal-state resistivity data fits the parallel resistor model described by Wiesmann \cite{ResistivityModel}. This model states that the temperature-dependent resistivity factor $\rho(T)$ consists of saturated resistivity at high temperatures $\rho_{s}$ and the ideal resistivity contribution $\rho_{ideal}(T)$. The mathematical expression for $\rho(T)$ is given as
\begin{equation}
\frac{1}{\rho(T)}= \frac{1}{\rho_{s}}+ \frac{1}{\rho_{ideal}(T)}.
\label{Eq1:Parallel}
\end{equation}
Here, $\rho_{ideal}(T)$ = $\rho_{i,0}$ + $\rho_{i,L}(T)$. $\rho_{i,0}$ is the temperature-independent residual resistivity resulting from impurity scattering, and $\rho_{i,L}(T)$ is the temperature-dependent resistivity resulting from electron-phonon scattering described by Wilson as \cite{Wilson},  
\begin{equation}
{\rho_{i,L}(T)}= A \left(\frac{T}{\Theta_{R}}\right)^n \int_{0}^{{\Theta_{R}}/T} \frac{x^{n}}{(e^{x}-1)(1-e^{-x})} dx.
\label{Eq2:Parallel}
\end{equation}\
Here, $A$ is a material-dependent quantity, $\Theta_{R}$ denotes the Debye temperature, and $n$ generally takes the values 2, 3, and 5 depending on the nature of the interaction \cite{el-phterm}. The best fit of the data for $n$ = 5 yielded Debye temperature $\Theta_{R}$ = 253(4) K, $\rho_{0}$ = 54.4(2) $\mu\Omega$ cm and $\rho_{sat}$ = 409.6(9) $\mu\Omega$ cm for ZrOs$_{2}$ and $\Theta_{R}$ = 240(3) K, $\rho_{0}$ = 87.0(7) $\mu\Omega$ cm and $\rho_{sat}$ = 176.0(3) $\mu\Omega$ cm for ZrOs$_{2}$, respectively, which are comparable to the C14 Re$_{2}$Hf compound \cite{Re2Hf}.

\begin{figure*}
\includegraphics[width=1.99\columnwidth]{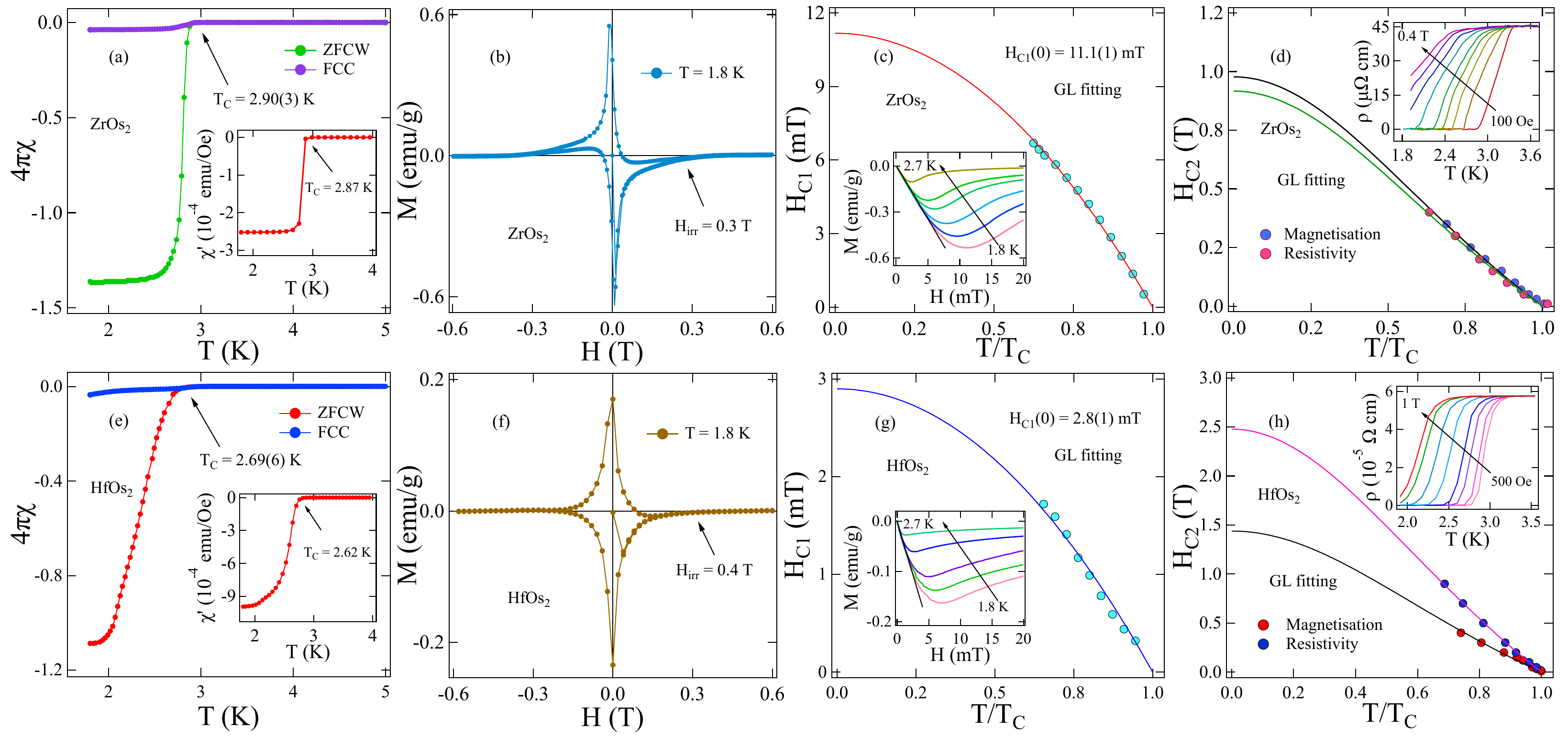}
\caption {\label{Fig3} (a) and (e) Temperature-dependent magnetic susceptibility measured in ZFCW and FCC mode at an applied magnetic field of 1 mT. The inset shows the real part of AC susceptibility. (b) and (f) Magnetization with variation of the magnetic field at 1.8 K. (c) and (g) Temperature dependence of the lower critical field, in which extrapolation of the G-L equation provides $H_{C1}$ (0); the inset shows the magnetic moment vs. the low field variation at different temperatures running from 1.8 to 2.7 K. (d) and (h) Temperature-dependent upper critical field, which is fitted using the G-L equation and its extrapolation provides H$_{C2}$(0). The inset shows magnetization with temperature variation at different applied magnetic fields.}
\end{figure*} 

\subsubsection{Magnetisation}
Temperature-dependent magnetization measurements were measured in the ZFCW (zero-field cooling) and FCC (field cooling) orders under an applied magnetic field of 1 mT for both compounds. The diamagnetic response that confirms the superconductivity at T$_{C}$ = 2.90(3) K for ZrOs$_{2}$ and 2.69(6) K for HfOs$_{2}$ as depicted in \figref{Fig3}(a) and (e). Strong magnetic-field pinning caused the separation of FCC from ZFCW, indicating type-II superconductivity with a full Meissner volume fraction in both samples. Superconductivity was further confirmed by testing AC susceptibility, as shown in the inset of \figref{Fig3}(a) and (e) for XOs$_{2}$ (X = Zr, Hf), respectively. Later, the type II nature of the superconductivity in both samples was confirmed by a magnetization loop under magnetic field variation at 1.8 K, which are shown in \figref{Fig3}(b) and (f). At 0.3 and 0.4 T, the irreversible magnetic fields, H$_{irr}$, are measured for the Zr and Hf variants, indicating the threshold over which vortices start to de-pin. 

Furthermore, to calculate the lower critical field H$_{C1}$(0), the magnetization variation in the low magnetic field at different isotherms from 1.8 K to the transition temperature was analyzed, as shown in the inset of \figref{Fig3}(c) and (g) for compounds based on Zr and Hf. H$_{C1}$ is the magnetic field value at which the curve deviates from the linear Meissner state for each isotherm. The temperature-dependent lower critical field is well-fitted with the Ginzburg-Landau (GL) equation described as follows:
\begin{equation}
H_{C1}(T)=H_{C1}(0)\left[1-\left(\frac{T}{T_{C}}\right)^{2}\right],
\label{eqn3:HC1}
\end{equation}
which gives H$_{C1}$(0) = 11.1(1) and 2.89(1) mT for ZrOs$_{2}$ and HfOs$_{2}$, respectively, as given in \figref{Fig3}(c) and (g). We measured magnetization and resistivity as the temperature changed in a number of different applied fields to look at the upper critical field H$_{C2}$(0). As the magnetic field increases, the resistivity curves in the inset of \figref{Fig3}(d) and (h) show a reduction in T$_{C}$, which is well characterized by the Ginzburg-Landau equation; 
\begin{equation}
H_{C2}(T) = H_{C2}(0)\left[\frac{1-t^{2}}{1+t^{2}}\right],  \quad  \text{where} \;  t = \frac{T}{T_{C}}
\label{eqn4:HC2}
\end{equation}
The obtained H$_{C2}$(0) values from GL fitting are 0.97(1) and 0.91(2) T for ZrOs$_{2}$ and 1.43(4) and 2.47(1) T for HfOs$_{2}$ from magnetization and resistivity measurements, respectively.

A magnetic field can destroy superconductivity via spin paramagnetic and orbital-limiting effects. According to the BCS theory, the spin paramagnetic effect results from pair breaking due to the Zeeman effect and is defined as the Pauli limiting field H$_{C2}^{P}$(0) = 1.86 T$_{C}$ \cite{Pauli1, Pauli2}. The obtained H$_{C2}^{P}$(0) values are 5.4(2) and 5.0(3) T, respectively, corresponding to observed T$_{C}$ values of 2.90(3) and 2.69(6) K for Zr and Hf variants. Furthermore, the orbital limiting effect increases the kinetic energy of one electron. It breaks the Cooper pair, which can be estimated by the Werthamer-Helfand-Hohenberg (WHH) theory for type-II superconductors while neglecting the spin-orbit coupling effect \cite{WHHM1, WHHM2}. The expression for an orbital limit of H$_{C2}$ is described by the following relation: 
\begin{equation}
H^{orb}_{C2}(0) = -\alpha T_{C} \left.{\frac{dH_{C2}(T)}{dT}}\right|_{T=T_{C}}, 
\label{eqn5:WHH}
\end{equation}
where constant $\alpha$ is the purity factor having values of 0.69 and 0.73 for dirty and clean limit superconductors, respectively. The initial slope $-\frac{dH_{C2}(T)}{dT}$ at T = T$_{C}$ is calculated to be 0.80 (1) and 0.66(4) T / K for the variants Zr and Hf. The estimated values of $H^{orb}_{C2}(0)$ for the variants Zr and Hf are 1.60 (1) and 1.21(5) T for $\alpha$ = 0.69 (for superconductors with dirty limits). The orbital effect may be responsible for the Cooper pair breaking, because both compounds' upper critical field values are substantially lower than the Pauli limiting field. The maki parameter (which is a measure of the strength of the Pauli limiting field and orbital critical field) is defined as $\alpha_{M}$ = $\sqrt{2}$ H$_{C2}^{orbital}$(0)/H$_{C2}^{P}$(0). The estimated values of $\alpha_{M}$ are 0.42(1) and 0.34(2) for the Zr and Hf compounds, respectively.
 
The Ginzburg-Landau coherence length $\xi_{GL}$ = 18.42(9) and 15.17(8) $nm$ for the variants Zr and Hf are estimated using the relation: $H_{C2}(0) = {\frac{\Phi_{0}}{2\pi \xi_{GL}^2(0)}}$. $\Phi_{0}$ is the magnetic flux quantum. For given Ginzburg-Landau coherence length and the lower critical field H$_{C1}(0)$ = 11.1(7) and 2.89(6) mT, the Ginzburg-Landau penetration depth $\lambda_{GL}$ are 190.47(5) and 445.64(3) $nm$ for variants Zr and Hf, calculated by given relation \cite{HC1, Tin}, 
\begin{equation}
H_{C1}(0) = \frac{\Phi_{0}}{4\pi\lambda_{GL}^2(0)}\left[ln \frac{\lambda_{GL}(0)}{\xi_{GL}(0)} + 0.12\right].
\label{eqn6:lamda}
\end{equation}
The GL parameter, which categorizes superconductors into type-I and type-II based on $\xi_{GL}$ and $\lambda_{GL}$, has the given formula $k_{GL}$ = $\frac{\lambda_{GL}(0)}{\xi_{GL}(0)}$. For the Zr and Hf variants, the obtained $k_{GL}$ values are 10.3(4) and 29.3(6), respectively, both exceeding $1/\sqrt{2}$. This indicates that the present systems are type-II superconductors. The thermodynamic critical field $H_{C}$ = 67.8(9) and 34.9(6) mT are calculated using the values of $H_{C2}$ and $H_{C1}$ with the help of a given relation \cite{Tin},
\begin{equation}
H_{C}^2 ln[k_{GL}(0)+0.08] = {H_{C1}(0) H_{C2}(0)}.
\label{eqn7:HC}
\end{equation}
All the superconducting parameters $\xi_{GL}$(0), $\lambda_{GL}$(0), $k_{GL}$, and $H_{C}$ are summarized in \tableref{tbl: parameters} for XOs$_{2}$.

\subsubsection{Specific heat}
\begin{figure}
\includegraphics[width=0.95\columnwidth]{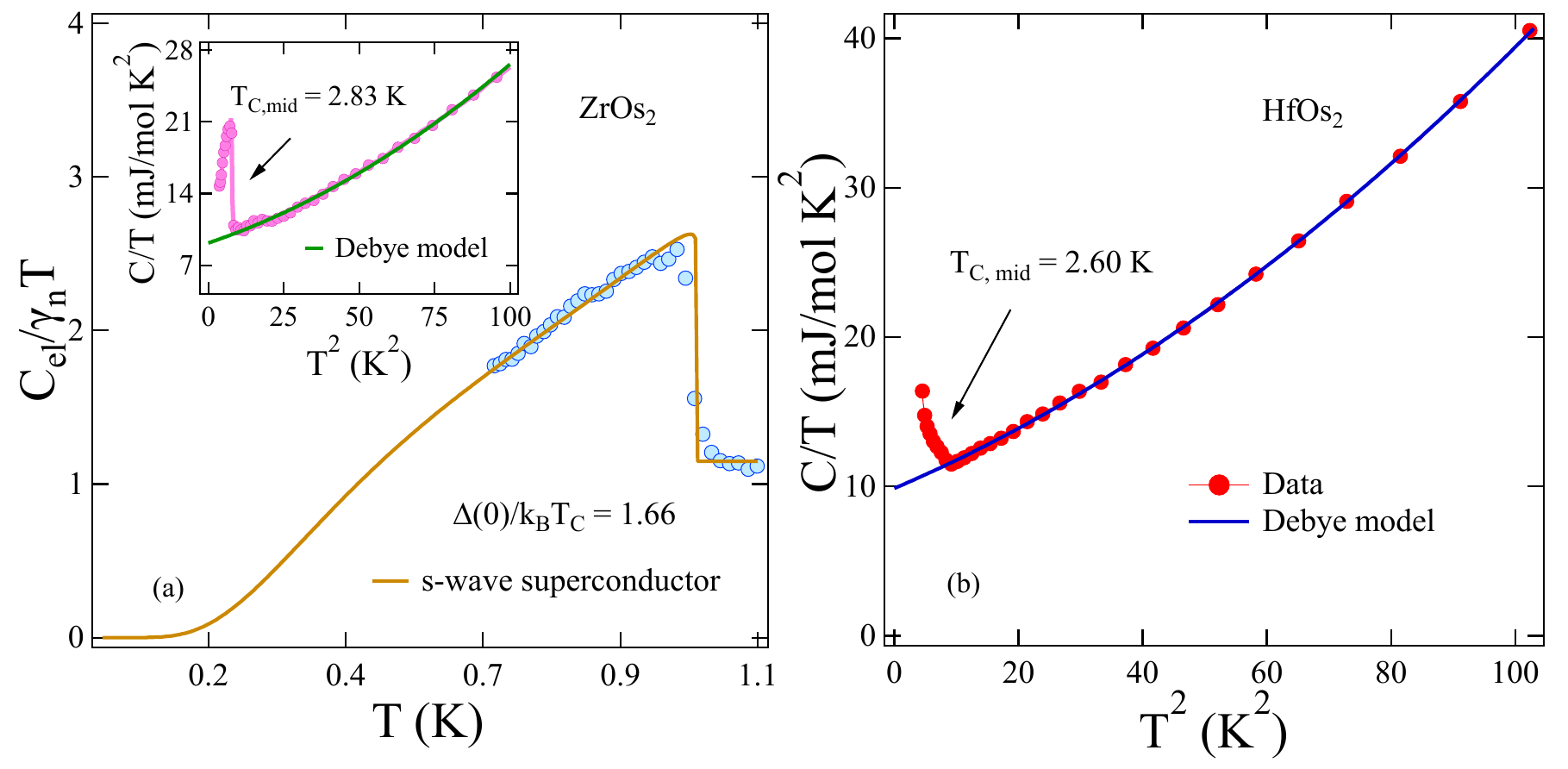}
\caption{\label{Fig4} (a) Specific heat data is well described by an isotropic s-wave model for ZrOs$_{2}$ shown by the solid brown line. Inset shows Specific heat versus temperature data which are well-fitted with the Debye relation and (b) Specific heat data of HfOs$_{2}$.}
\end{figure}

The insets of \figref{Fig4}(a) and \figref{Fig4}(b) show the temperature-dependent specific heat in a zero magnetic field (H = 0 mT) for both compounds. A significant jump in specific heat occurs at transition temperatures T$_{C}$ = 2.8 and 2.6 K for ZrOs$_{2}$ and HfOs$_{2}$, which is comparable to the T$_{C}$ observed from the other measurements. In the normal state, the data were fitted using the Debye relation given as $C$ = $\gamma_{n}T$ + $\beta_{3}T^{3}$ + $\beta_{5}T^{5}$, where the electronic contribution to the specific heat is represented by $\gamma_{n}T$, the phononic contribution by $\beta_{3}T^{3}$ and the anharmonic contribution by $\beta_{5}T^{5}$. The best fit to the data provides the Sommerfeld coefficient $\gamma_{n}$ = 8.02(1) mJ/mol K$^{2}$, Debye constant, $\beta_{3}$ = 0.14(3) mJ/mol K$^{4}$ and $\beta_{5}$ = 0.00041(3) mJ/mol K$^{6}$ for ZrOs$_{2}$ and $\gamma_{n}$ = 9.82(4) mJ/mol K$^{2}$, $\beta_{3}$ = 0.18(1) mJ/mol K$^{4}$ and $\beta_{5}$ = 0.0011(5) mJ/mol K$^{6}$ for HfOs$_{2}$, respectively.

Furthermore, the density of states at the Fermi level $D_{C}(E_{F})$ can be calculated using the relation given as $\gamma_{n}$ = $\left(\frac{\pi^{2} k_{B}^{2}}{3}\right) D_{C}(E_{F})$, where $k_{B}$ = 1.38 $\times$ 10$^{-23}$ J K$^{-1}$. The resulting values of $D_{C}(E_{F})$ are 3.40(1) and 4.16(5) states eV$^{-1}$f.u.$^{-1}$ for ZrOs$_{2}$ and HfOs$_{2}$, comparable to the calculated values for Ru-Based Laves phase \cite{Ru-LavesPhase}. The Debye temperature $\theta_{D}$, that is, the temperature of the highest normal mode of vibration, is also obtained using the Debye constant $\beta_{3}$ by the given expression, 
\begin{equation}
\theta_{D} = \left(\frac{12\pi^{4} R N}{5 \beta_{3}}\right)^{\frac{1}{3}},
\label{eqn8:DebyeTemperature}
\end{equation}
where R = 8.314 J mol$^{-1}$ K$^{-1}$ is a gas constant. For $N$ = 3 (XOs$_{2}$) number of atoms per formula unit, the estimated values of $\theta_{D}$ are 346 (2) and 318 (3) K for the Zr and Hf compounds, consistent with the theoretically calculated elastic modules \cite{TMOs2}. Electron-phonon coupling $\lambda_{e-ph}$, a dimensionless number that assesses the strength of the attractive interaction between electrons and phonons, can be calculated using the value of T$_{C}$ and $\theta_{D}$ from McMillan's equation \cite{el-ph},
\begin{equation}
\lambda_{e-ph} = \frac{1.04+\mu^{*}\mathrm{ln}(\theta_{D}/1.45T_{C})}{(1-0.62\mu^{*})\mathrm{ln}(\theta_{D}/1.45T_{C})-1.04 };
\label{eqn9:Lambda}
\end{equation}
Here, $\mu^{*}$ is a repulsively screened Coulomb pseudopotential parameter taken as 0.13 for intermetallic compounds. The calculated $\lambda_{e-ph}$ values are 0.53(4) and 0.53(6) for ZrOs$_{2}$ and HfOs$_{2}$, for given values of T$_{C}$ = 2.90(3) and 2.69(6) K and $\theta_{D}$ = 346 (2) and 318 (3) K, suggesting weakly coupled BCS superconductivity in both compounds. The parameters obtained $D_{C}(E_{F})$ and $\lambda_{e-ph}$ are comparable to those of other superconducting Laves phase compounds \cite{RRu2}.

Specific heat data also help to understand more about the superconducting gap's symmetry and properties. Electronic specific heat at zero applied field has been estimated by subtracting the phononic contribution from total specific heat by the expression C$_{el} = C - \beta_{3}T^{3} - \beta_{5}T^{5}$ for ZrOs$_{2}$; however, electronic specific heat calculation for HfOs$_{2}$ is not shown due to insufficient data points at low temperature in their superconducting state. The low-temperature electronic specific heat plot in \figref{Fig4}(a) for ZrOs$_{2}$ fits well with the $\alpha$ model of the single-gap BCS superconductor \cite{BCSModel}. According to this model, the entropy $S$ can be expressed as,
\begin{equation}
\frac{S}{\gamma_{n} T_{C}}= -\frac{6}{\pi^{2}} \left(\frac{\Delta(0)}{k_{B} T_{C}}\right) \int_{0}^{\infty} \left[ {fln(f)+(1-f)ln(1-f)} \right] dy
\label{eqn10:BCS}
\end{equation}
Here, $f(\xi)$ = $[e^{\beta E(\xi)}+1]^{-1}$; with $\beta = (k_{B}T)^{-1}$ being the Fermi function, whereas the integration variable is $y$ = $\xi/\Delta(0)$. E($\xi$) is expressed as $\sqrt{\xi^{2}+\Delta^{2}(t)}$ which represents the energy of normal electrons relative to the Fermi energy, and $\Delta(t)$ is the temperature-dependent gap function with reduced temperature, t = $T/T_{C}$. The gap function can be written as $\Delta(t) = tanh[1.82(1.018(1/t))-1]^{0.51}$ in the case of an isotropic s-wave BCS approximation. The electronic specific heat, C$_{el}$, can be related to $S$ via the relation C$_{el}$ = $tdS/dt$. The best fit of the data using the isotropic s-wave BCS model, as shown in \figref{Fig4}(a), yields a superconducting gap value $\Delta(0)/k_{B}T_{C}$ of 1.66 (1) for ZrOs$_{2}$, which is close to the standard BCS gap value in the weak coupling limit.

\subsection{Electronic parameters and Uemura plot}
A series of equations were used to calculate electronic parameters such as the mean free path $l_{e}$, the effective mass $m^{*}$, and the coherence length considering the spherical Fermi surface \cite{5equations}. The equations are given as:
\begin{equation}
\gamma_{n} = \left(\frac{\pi}{3}\right)^{2/3}\frac{k_{B}^{2}m^{*}V_{\mathrm{f.u.}}n^{1/3}}{\hbar^{2}N_{A}},
\label{eqn11:gf}
\end{equation}

\begin{equation}
\textit{l}_{e} = \frac{3\pi^{2}{\hbar}^{3}}{e^{2}\rho_{0}m^{*2}v_{\mathrm{F}}^{2}}, n = \frac{1}{3\pi^{2}}\left(\frac{m^{*}v_{\mathrm{F}}}{\hbar}\right)^{3}.
\label{eqn12:le,n}
\end{equation}
where $k_{B}$ is the Boltzmann constant, $N_{A}$ is the Avogadro number, and $V_{f.u.}$ is the volume of formula unit. Furthermore, the GL penetration depth $\lambda_{GL}$(0) is affected in the case of dirty limit superconductors, which can be expressed in terms of London penetration depth $\lambda_{L}$ by the given expression,
\begin{equation}
\lambda_{GL}(0) = 
\lambda_{L}
\left(1+\frac{\xi_{0}}{\textit{l}_{e}}\right)^{1/2}, \lambda_{L} = 
\left(\frac{m^*}{\mu_0
ne^{2}}\right)^{1/2}
\label{eqn13:f}
\end{equation}
where $\xi_{0}$ is the BCS coherence length, which is also expressed in terms of the GL coherence length $\xi_{GL}$ by the given expression, 
\begin{equation}
\frac{\xi_{GL}(0)}{\xi_{0}} = \frac{\pi}{2\sqrt{3}}\left(1+\frac{\xi_{0}}{\textit{l}_{e}}\right)^{-1/2}.
\label{eqn14:xil}
\end{equation}

\begin{figure}
\includegraphics[width=0.95\columnwidth]{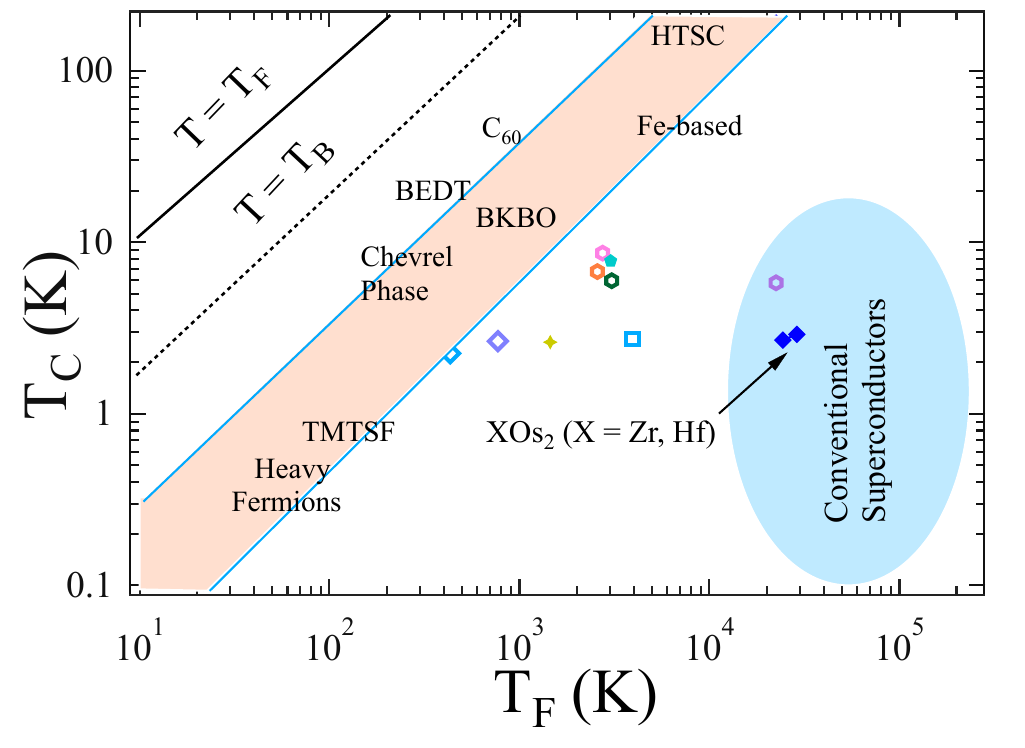}
\caption {\label{Fig5} The Uemura plot classifies the superconductor in a conventional and unconventional form based on critical and Fermi temperatures. XOs$_{2}$ (X = Zr, Hf) is positioned as a blue marker near the conventional range of superconductivity.}
\end{figure} 

\begin{table}
\caption{Superconducting and normal state parameters of ZrOs$_{2}$ and HfOs$_{2}$, derived from magnetization, resistivity, and specific heat measurement compared with HfRe$_{2}$ \cite{Re2Hf}.}
\label{tbl: parameters}
\setlength{\tabcolsep}{3pt}
\begin{center}
\begin{tabular}[b]{ l c c c c}\hline \hline
Parameters& unit& ZrOs$_{2}$ & HfOs$_{2}$ & HfRe$_{2}$ \\
\hline
%\\[0.1ex]                           
T$_{C}$ & K &2.90(3) &2.69(6) & 5.7(1)\\
H$_{C1}(0)$ & mT & 11.1(1) &2.8(1) &12.7(1)\\ 
H$_{C2}^{mag}$(0) & T & 0.97(1) &1.43(4)& 1.17(2) \\
H$_{C2}^{P}$(0) & T & 5.4(2) &5.0(3)&10.6 \\
$\xi_{GL}$& $nm$& 18.42(9) &15.17(8)&16.7 \\
$\lambda_{GL}^{mag}$& $nm$& 190.47(5)&445.64(3) &174.2\\
$k_{GL}$& - & 10.3(4) &29.3(6)&10.38\\
$\gamma_{n}$& mJ/mol K$^{2}$&  8.02(1) &9.82(4)& 11.1\\
$\theta_{D}$& K&346(2)&318(3) &294\\
$\frac{\Delta(0)}{k_{B}T_{C}}$ (sp) &  - & 1.66(1) & - &1.80\\
$\lambda_{e-ph}$ & - & 0.53(4) &0.53(6)&0.67\\
$D_{C}(E_{F})$ & states/(eV f.u.) & 3.40(1) &4.16(5)& -\\ 
$m^{*}/m_{e}$& - &0.8(1)&0.97(2)&-\\
$n$&10$^{28}$&1.25(5)&1.29(3)&-\\
V$_{f}$&10$^{5}$ m/s&10.4(1)&8.6(7)&-\\
$\xi_{0}$&10$^{-10}$ m &8.3(8)&37.6(2)&-\\
$l$&10$^{-11}$ m&4.3(3)&4.0(2)&-\\
T$_{F}$& K &28311&24294&-\\
%\\[0.1ex]
\hline \hline
\end{tabular}
\par\medskip\footnotesize
\end{center}
\end{table}

The calculated values of all these electronic parameters are mentioned in \tableref{tbl: parameters}. Furthermore, the Fermi temperature can be calculated using the given relation, where $n$ and $m^{*}$ are the electronic and effective mass of quasi-particles, respectively \cite{Tf}:
\begin{equation}
k_{B}T_{F} = \frac{\hbar^{2}}{2}(3\pi^{2})^{2/3}\frac{n^{2/3}}{m^{*}}.
\label{eqn15:tf}
\end{equation}
Based on the ratio of T$_{C}$ and T$_{F}$, the Uemura scheme classifies superconductors as conventional or unconventional superconductors \cite{CS/US}. The unconventional range of the superconductor is shown in \figref{Fig5} by the band between 0.01 $\leq$ T$_{C}$/T$_{F}$ $\leq$ 0.1. Compounds having T$_{C}$/T$_{F}$ $\geq$ 0.1 are considered conventional superconductors, while exotic superconductors such as high T$_{C}$ cuprates, heavy fermions, Chervral phases, and organic superconductors are classified as unconventional. The compounds are within the conventional BCS range, given the calculated values of $T_{F}$ = 28877 and 24294 K and the observed values of $T_{C}$ = 2.90 (3) and 2.69 (6) K for the variants Zr and Hf, respectively. The blue marker shows their positions on the Uemura plot in \figref{Fig5}.

\section{Conclusion}
Laves phase XOs$_{2}$ (X = Zr, Hf) were synthesized and the superconducting properties were investigated using electrical transport, magnetization, and specific heat experiments. XOs$_{2}$ adopts a hexagonal MgZn$_{2}$-type structure with space group P6$_{3}$/$mmc$, referred to as a breathing kagome compound. The metallic nature of the samples is confirmed by resistivity. Magnetization measurement confirms bulk type-II superconductivity with a transition temperature T$_{C}$ = 2.90(3) and 2.69(6) K for ZrOs$_{2}$ and HfOs$_{2}$. All physical parameters, as detailed in \tableref{tbl: parameters}, are consistent with those observed in the candidate time-reversal symmetry breaking (TRSB) HfRe$_{2}$ C14 Laves phase and the recent findings on ROs$_{2}$, where R spans from Sc, Y and Lu in a breathing kagome compound \cite{Re2Hf, ROs2}. Specific heat data in a zero magnetic field reveal a jump value of 1.66, indicative of fully gapped superconductivity in ZrOs$_{2}$. To further unravel the exact pairing mechanism in these compounds, comprehensive microscopic measurements, such as muon spin relaxation and rotation, on Laves phase-based kagome single and polycrystalline samples will be crucial, as this research has the potential to significantly advance our understanding of the superconducting ground state of the AV$_{3}$Sb$_{5}$ compounds, which is challenging due to the intricate interplay between charge density waves and superconductivity.

\section{Acknowledgments} 
P.~K.~M. acknowledges the funding agency Council of Scientific and Industrial Research, Government of India, for providing the SRF fellowship (Award No: 09/1020(0174)/2019-EMR-I). R.~P.~S. acknowledges the Science and Engineering Research Board, Government of India, for the Core Research Grant CRG/2019/001028.


\begin{thebibliography}{50}
\bibitem{Hasan} J. X. Yin, B. Lian, and M. Z. Hasan, Nature 612, 647 (2022).
\bibitem{Review} Y. Wang, H. Wu, G. T. McCandless, J. Y. Chan, and M. N. Ali, arXiv:2303.03359 (2023).
\bibitem{magneticfrustration1} S. Yan, D. A. Huse, and S. R. White, Science 332, 1173 (2011).
\bibitem{magneticfrustration2} T.H. Han, J. S. Helton, S. Chu, D. G. Nocera, J. A. Rodriguez-Rivera, C. Broholm, and Y. S. Lee, Nature (London) 492, 406 (2012).
\bibitem{Co3Sn2S2-1} D. F. Liu, A. J. Liang, E. K. Liu, Q. N. Xu, Y. W. Li, C. Chen, D. Pei, W. J. Shi, S. K. Mo, P. Dudin, and T. Kim, Science 365, 1282 (2019).
\bibitem{GFS1} S. T. Bramwell and M. J. P. Gingras, Science 294, 1495 (2001).
\bibitem{GFS2} S. H. Lee, C. Broholm, W. Ratcliff, G. Gasparovic, Q. Huang, T. H. Kim, and S. W. Cheong, Nature (London) 418, 856 (2002).
\bibitem{AV3Sb5-1} B. R. Ortiz, S. M. L. Teicher, Y. Hu, J. L. Zuo, P. M. Sarte, E. C. Schueller, A. M. M. Abeykoon, M. J. Krogstad, S. Rosenkranz, R. Osborn, R. Seshadri, L. Balents, J. He, and S. D. Wilson, Phys. Rev. Lett. 125, 247002 (2020).
\bibitem{AV3Sb5-2} B. R. Ortiz, P. M. Sarte, E. M. Kenney, M. J. Graf, S. M. L. Teicher, R. Seshadri, and S. D. Wilson, Phys. Rev. Mater. 5, 034801 (2021).
\bibitem{AV3Sb5-3} C. Mielke III, D. Das, J.-X. Yin, H. Liu, R. Gupta, C. N. Wang, Y.-X. Jiang, M. Medarde, X. Wu, H. C. Lei, J. J. Chang, P. Dai, Q. Si, H. Miao, R. Thomale, T. Neupert, Y. Shi, R. Khasanov, M. Z. Hasan, H. Luetkens, and Z. Guguchia, Nature (London) 602, 245 (2022).
\bibitem{Kagome-nontrivial} Z. Li, J. Zhuang, L. Wang, H. Feng, Q. Gao, X. Xu, W. Hao, X. Wang, C. Zhang, K. Wu, S. X. Dou, L. Chen, Z. Hu, and Y. Du, Sci. Adv. 4, 1343 (2018).
\bibitem{YMn6Sn6} M. Li, Q. Wang, G. Wang, Z. Yuan, W. Song, R. Lou, Z. Lu, Y. Huang, Z. Liu, H. Lei, Z. Yin, and S. Wang, Nat. Commun. 12, 3129 (2021).
\bibitem{Flatband1} Z. Sun, H. Zhou, C. Wang, S. Kumar, D. Geng, S. Yue, X. Han, Y. Haraguchi, K. Shimada, P. Cheng, L. Chen, Y. Shi, K. Wu, S. Meng, and B. Feng, Nano Lett. 22, 4596 (2022).
\bibitem{Flatband2} M. Kang, S. Fang, L. Ye, H. C. Po, J. Denlinger, C. Jozwiak, A. Bostwick, E. Rotenberg, E. Kaxiras, J. G. Checkelsky, and R. Comin, Nat. Commun. 11, 4004 (2020).
\bibitem{Flatband3} E. Uykur, B. R. Ortiz, S. D. Wilson, M. Dressel, and A. A. Tsirlin, NPJ Quantum Mater. 7, 16 (2022).
\bibitem{Flatband4} H. Zhang, Z. Shi, Z. Jiang, M. Yang, J. Zhang, Z. Meng, T. Hu, F. Liu, L. Cheng, Y. Xie, and J. Zhuang, Adv. Mater. 2301790 (2023).
\bibitem{BreathingKagome-nontrivial1} A. Bolens and Nagaosa, Phys. Rev. B 99, 165141 (2019).
\bibitem{BreathingKagome-nontrivial2} H. Tanaka, Y. Fujisawa, K. Kuroda, R. Noguchi, S. Sakuragi, C. Bareille, B. Smith, C. Cacho, S. W. Jung, T. Muro, Y. Okada, and T. Kondo, Phys. Rev. B 101, 161114(R) (2020). 
\bibitem{BreathingKagome-nontrivial3} Z. Sun, H. Zhou, C. Wang, S. Kumar, D. Geng, S. Yue, X. Han, Y. Haraguchi, K. Shimada, P. Cheng, L. Chen, Y. Shi, K. Wu, S. Meng, and B. Feng, Nano Lett. 22, 4596 (2022). 
\bibitem{BreathingKagome-nontrivial4} S. Regmi, T. Fernando, Y. Zhao, A. P. Sakhya, G. Dhakal, I. Bin Elius, H. Vazquez, J. D. Denlinger, J. Yang, J.-H. Chu, X. Xu, T. Cao, and M. Neupane, Commun. Mater. 3, 100 (2022).
\bibitem{Kagome-muon1} Z. Guguchia, R. Khasanov, and H. Luetkens, npj Quantum Materials 8, 41 (2023).
\bibitem{Kagome-muon2} Z. Shan, P. K. Biswas, S. K. Ghosh, T. Tula, A. D. Hillier, D. Adroja, S. Cottrell, G. H. Cao, Y. Liu, X. Xu, Y. Song, H. Yuan, and M. Smidman, Phys. Rev. Research 4, 033145 (2022).
\bibitem{Kagome-Review} K. Jiang, T. Wu, J.-X. Yin, Z. Wang, M. Zahid Hasan, S. D. Wilson, X. Chen, and J. Hu, National Sci. Rev. 10, nwac199 (2023).
\bibitem{Mg2Ir3Si} K. Kudo, H. Hiiragi, T. Honda, K. Fujimura, H. Idei, and M. Nohara, J. Phys. Soc. Jpn. 89, 013701 (2020).
\bibitem{BreathingKagome} H. Liu, J. Yao, J. Shi, Z. Yang, D. Yan, Y. Li,  D. Chen, H. L. Feng, S. Li, Z. Wang, and Y. Shi, arXiv:2306.03370 (2023).
\bibitem{ROs2} K. Górnicka, M. J. Winiarski, D. I. Walicka, and T. Klimczuk, Scientific Reports 13, 16704 (2023).
\bibitem{Re2Hf} M. Mandal, A. Kataria, C. Patra, D. Singh, P. K. Biswas, A. D. Hillier, T. Das, and R. P. Singh, Phys. Rev. B 105, 094513 (2022).
\bibitem{LaRu3Si2} C. Mielke III, Y. Qin, J.-X. Yin, H. Nakamura, D. Das, K. Guo, R. Khasanov, J. Chang, Z. Q. Wang, S. Jia, S. Nakatsuji, A. Amato, H. Luetkens, G. Xu, M. Z. Hasan, and Z. Guguchia, Phys. Rev. Mater. 5, 034803 (2021).
\bibitem{LaIr3Ga2} X. Gui, and R. J. Cava, Chem. Mater. 34, 2824 (2022).
\bibitem{YRu3Si2} C. S. Gong, S. J. Tian, Z. J. Tu, Q. W. Yin, Y. Fu, R. T. Luo, and H. C. Lei, Chin. Phys. Lett. 39, 087401 (2022).
\bibitem{LaRh3B2} S. Chaudhary, Shama, J. Singh, A. Consiglio, D. Di Sante, R. Thomale, and Y. Singh, Phys. Rev. B 107, 085103 (2023).
\bibitem{BaPt3B2} S. Li, J. Xing, J. Tao, H. Yang, and H. H. Wen, Ann. Phys. 358, 248 (2015).
\bibitem{CeOs3B2} K. S. Athreya, L. S. Hausermannberg, R. N. Shelton, S. K. Malik, A. M. Umarji, and G. K. Shenoy, Phys. Lett. A 113, 330 (1985).
\bibitem{Laves-Phase1} R. L. Berry and G. V. Raynor, Acta Cryst. 6, 178 (1953).
\bibitem{electronegativity-1} A. Ormeci, A. Simon, and Y. Grin, Angew. Chem. Int. Ed. Engl. 49, 8997 (2010).
\bibitem{1} F. Stein, M. Palm, and G. Sauthoff, Intermetallics 12, 713 (2004).
\bibitem{Frank-Kasper} F. C. Frank and J. S. Kasper, Acta Cryst. 11, 184 (1958).
\bibitem{Non-symmorphic1} H. Watanabe, H. C. Po, A. Vishwanath, and M. Zaletel, Proc. Natl. Acad. Sci. U.S.A. 112, 14551 (2015).
\bibitem{TSM} B. Bradlyn, L. Elcoro, J. Cano, M. G. Vergniory, Z. Wang, C. Felser, M. I. Aroyo, and B. A. Bernevig, Nature (London) 547, 298 (2017).
\bibitem{Theortical-ZrOs2} Q. J. Liu, N. C. Zhang, F. S. Liu, and Z. T. Liu, Journal of alloys and compounds 589, 278 (2014).
\bibitem{Hf-DFTcalculation} V. S. Sathyakumari, S. Sankar, and K. Mahalakshmi, Sci-Poland 32(2), 324 (2014).
\bibitem{Os2Zr} R. Kuentzler and R. W. Waterstrat, J. Less-Common Met. 125, 261 (1986).
\bibitem{Pr-ZrOs2} A. Slebarski, D. Wohllebenand, and P. Weidner, Z. Phys. B: Condens. Matter 61, 177 (1985).
\bibitem{Topological-SC} M. Sato, and Y. Ando, Rep. Prog. Phys. 80, 076501 (2017).
\bibitem{Unconventional-SCs} M. Sigrist and K. Ueda, Rev. Mod. Phys. 63, 239 (1991).
\bibitem{VESTA} K. Momma, and F. Izumi, J. Appl. Crystallogr. 44, 1272 (2011).
\bibitem{Lavesphase6} R. L. Johnston and R. Hoffmann, Z. Anorg. Allg. Chem. 616, 105 (1992).
\bibitem{Laves-phase2} V. B. Compton, and B. T. Matthias, Acta Crystallogr A 12, 651 (1959).
\bibitem{FullProf} J. Rodriguez-Carvajal, Physica B. 192, 55 (1993).
\bibitem{ZrOs2} S. L. McCarthy and L. Schmidt, J. Less-Common Met. 23, 241 (1971).
\bibitem{atomicradius} J. C. Slater, J. Chem. Phys. 41, 3199 (1964).
\bibitem{LS1} B. T. Matthias, V. B. Compton, and E. Corenzwit, J. Phys. Chem. Solids 19, 130 (1961).
\bibitem{Ir-LavesPhase} T. H. Geballe, B. T. Matthias, V. B. Compton, E. Corenzwit, G. W. Hull, Jr., and L. D. Longinotti, Phys. Rev. 137, A119 (1965).
\bibitem{Ru-LavesPhase} S. Niitaka, E. Minamitani, Y. Kim, H. Takagi, and K. Kono, J. Phys. Soc. Jpn. 82, 124703 (2013).
\bibitem{ResistivityModel} H. Wiesmann, M. Gurvitch, H. Lutz, A. K. Ghosh, B. Schwarz, M. Strongin, P. B. Allen, and J. W. Halley, Phys. Rev. Lett. 38, 782 (1977).
\bibitem{Wilson} G. Grimvall, The Electron-Phonon Interaction in Metals (North-Holland, Amsterdam, 1981).
\bibitem{el-phterm} A. Bid, A. Bora, and A. K. Raychaudhuri, Phys. Rev. B 74, 035426 (2006).
\bibitem{Pauli1} B. S. Chandrasekhar, Appl. Phys. Lett. 1, 7 (1962).
\bibitem{Pauli2} A. M. Clogston, Phys. Rev. Lett. 9, 266 (1962).
\bibitem{WHHM1} N. R. Werthamer, E. Helfand, and P. C. Hohenberg, Phys. Rev. 147, 295 (1966).
\bibitem{WHHM2} E. Helfand, and N. R. Werthamer, Phys. Rev. 147, 288 (1966).
\bibitem{HC1} T. Klimczuk, F. Ronning, V. Sidorov, R. J. Cava, and J. D. Thompson, Phys. Rev. Lett. 99, 257004 (2007).
\bibitem{Tin} M. Tinkham, Introduction to Superconductivity, 2nd ed. (McGraw-Hill, New York, 1996).
\bibitem{TMOs2} D. Qu, L. Bao, Z. Kong, and Y. Duan, Mater. Res. Express 6, 116569 (2019).
\bibitem{el-ph} W. L. McMillan, Phys. Rev. 167, 331 (1968).
\bibitem{RRu2} Niitaka, E. Minamitani, Y. Kim, H. Takagi, and K. Kono, J. Phys.Soc. Jpn.82, 124703 (2013).
\bibitem{BCSModel} H. Padamsee, J. E. Neighbor, and C. A. Shiffman, J. Low Temp. Phys. 12, 387 (1973).
\bibitem{5equations} D. A. Mayoh, J. A. T. Barker, R. P. Singh, G. Balakrishnan, D. McK. Paul, and M. R. Lees, Phys. Rev. B 96, 064521 (2017).
\bibitem{Tf} A. D. Hillier and R. Cywinski, Appl. Magn. Reson. 13, 95 (1997).
\bibitem{CS/US} Y. J. Uemura, G. M. Luke, B. J. Sternlieb, J. H. Brewer, J. F. Carolan, W. N. Hardy, R. Kadono, J. R. Kempton, R. F. Kiefl, S. R. Kreitzman, P. Mulhern, T. M. Riseman, D. L. Williams, B. X. Yang, S. Uchida, H. Takagi, J. Gopalakrishnan, A. W. Sleight, M. A. Subramanian, C. L. Chien, M. Z. Cieplak, G. Xiao, V. Y. Lee, B. W. Statt, C. E. Stronach, W. J. Kossler, and X. H. Yu, Phys. Rev. Lett. 62, 2317 (1989).
\end{thebibliography}
\end{document}